%% file: main.tex
\documentclass[twocolumn,english,aps,prb,superscriptaddress,bibnotes,amsmath,amssymb,floatfix,nobalancelastpage]{revtex4-2}
\usepackage[colorlinks=true,citecolor=blue,linkcolor=magenta]{hyperref}
\usepackage[utf8]{inputenc}
\usepackage[english]{babel}
\usepackage{amsmath,amsfonts,amssymb}
\usepackage[T1]{fontenc}
\usepackage{textcomp}
\usepackage{url}
\usepackage{soul}
\usepackage{changes}
\usepackage{amsmath}
\usepackage{amsfonts}
\usepackage{amssymb}
\usepackage[version=3]{mhchem}
\usepackage{stmaryrd}
\usepackage{epstopdf}
\usepackage{graphicx}
\usepackage{lettrine}
\usepackage{lipsum}
\usepackage{tabularray}
\newcommand{\Fref}[1]{Figure~\ref{#1}}
\newcommand{\fref}[1]{Fig.~\ref{#1}}

\begin{document}

\title{Brillouin-Enhanced Photonic Stepped-Frequency Radar}
%--------------------------1---------------------------
\author{Ziqian Zhang}
\email{ziqian.zhang@sydney.edu.au}
\affiliation{Institute of Photonics and Optical Science (IPOS), School of Physics, The University of Sydney, NSW 2006, Australia}
\affiliation{The University of Sydney Nano Institute (Sydney Nano), The University of Sydney, NSW 2006, Australia}
\affiliation{ARC Centre of Excellence in Optical Microcombs for Breakthrough Science (COMBS), NSW 2006, Australia}
%--------------------------2---------------------------
\author{Ryan L. Russell}
\affiliation{Institute of Photonics and Optical Science (IPOS), School of Physics, The University of Sydney, NSW 2006, Australia}
\affiliation{The University of Sydney Nano Institute (Sydney Nano), The University of Sydney, NSW 2006, Australia}
\affiliation{ARC Centre of Excellence in Optical Microcombs for Breakthrough Science (COMBS), NSW 2006, Australia}
%--------------------------3---------------------------
\author{Choon Kong Lai}
\affiliation{Institute of Photonics and Optical Science (IPOS), School of Physics, The University of Sydney, NSW 2006, Australia}
\affiliation{The University of Sydney Nano Institute (Sydney Nano), The University of Sydney, NSW 2006, Australia}
\affiliation{ARC Centre of Excellence in Optical Microcombs for Breakthrough Science (COMBS), NSW 2006, Australia}
%--------------------------5---------------------------
\author{Benjamin J. Eggleton}
\email{benjamin.eggleton@sydney.edu.au}
\affiliation{Institute of Photonics and Optical Science (IPOS), School of Physics, The University of Sydney, NSW 2006, Australia}
\affiliation{The University of Sydney Nano Institute (Sydney Nano), The University of Sydney, NSW 2006, Australia}
\affiliation{ARC Centre of Excellence in Optical Microcombs for Breakthrough Science (COMBS), NSW 2006, Australia}

% \date{\today}
\maketitle
%------------------------abstract----------------------
% abstract 
\noindent \textbf{Photonic stepped-frequency (SF) radar offers high range resolution and only requires low-speed driving electronics, but existing architectures face challenges in achieving low phase noise and uniform frequency steps simultaneously. Here, we demonstrate a photonic SF radar system that exploits dual Brillouin lasers in a shared fiber cavity to simultaneously suppress phase noise and ensure uniform frequency stepping. Phase noise is reduced through Brillouin optomechanical suppression and common-mode noise rejection upon photomixing. Frequency-step uniformity is enforced via lasing at a series of uniformly spaced cavity resonances. The system generates an X-band SF waveform spanning 1.31 GHz, achieving >23 dB of phase-noise improvement at a 100 kHz offset relative to a low-cost driving voltage-controlled oscillator. The demonstrated system reduces the dependence of the output waveform quality on noise in the driving electronics, offering a path towards high-performance radar sensing.}

%------------------------introduction------------------
\noindent Radar systems are fundamental to modern sensing, driving applications from indoor vital-sign monitoring \cite{zhang2023photonic} to satellite-based exploration of Mars' ionosphere \cite{orosei2018radar}. To meet evolving sensing demands, microwave photonics (MWP) \cite{yao2022microwave, marpaung2019integrated} offers a pathway to next-generation performance by enabling low-noise \cite{ghelfi2014fully} signal generation at high carrier frequencies \cite{kittlaus2021low}. By accessing these higher frequencies, systems can utilize broader RF bandwidths to achieve superior range resolution \cite{pan2020microwave}. Furthermore, the low transmission loss of photonic links enables distributed, multi-band sensing \cite{ghelfi2016photonics, serafino2019toward, huang2019centralized} and radar-LiDAR fusion \cite{falconi2020combined}, further improving overall system robustness.

Radar’s range resolution improves with increasing bandwidth, allowing the system to separate closer targets. Photonic radars synthesize a wide bandwidth through frequency modulation, using either linear (LFM) or stepped (SF) waveforms. While LFM requires a continuous wideband sweep within each pulse, SF modulation synthesizes an equivalent bandwidth through discrete, narrowband steps \cite{Ozdemir2021}. This approach lowers the instantaneous bandwidth requirements for the electronics driving the photonic generator and receiver \cite{Einstein1984, Nguyen2016}. As a result, photonic SF radar achieves high-resolution sensing with low-speed, MHz-level electronics, alleviating the need for costly high-performance components and enabling the practical deployment of photonic radar.

To date, photonic SF radar systems employ three main approaches for waveform generation, each with distinct limitations. The first approach employs a photonics-based RF multiplication technique \cite{pinna2016photonics, melo2018photonics, zhang2017photonics}, which multiplies an RF waveform from an electronic function generator and outputs bandwidth-broadened SF. The multiplication process not only increases the bandwidth of the input RF signal but also amplifies its phase noise, imposing a noise penalty that reduces the output signal quality (see Supplementary). The second approach uses frequency-shifting loops, where noise accumulates with each round-trip in the loop and progressively degrades signal-to-noise ratio (SNR) \cite{zhang2020multioctave, fernandez2022fundamental, zhang2025theoretical, guan2025temporal, ma2023coherent, ma2022high}, creating a fundamental trade-off between bandwidth and noise. The third approach uses laser wavelength sweeping—via direct current modulation or injection locking in a master-slave laser configuration—but nonlinear tuning dynamics introduce time-frequency nonlinearity \cite{zhou2016flexible, zhou2017reconfigurable, yu2024broadband, li2024stepped}, resulting in inconsistent frequency steps that degrade radar sensing performance. Therefore, a photonic approach that simultaneously achieves low phase noise, wideband operation, and high time-frequency linearity without relying on high-speed electronics or complex system architecture remains a key goal. 

In this work, we demonstrate a low-noise photonic SF radar enabled by dual Brillouin lasers (BLs) in a high-quality (high-Q) factor fiber ring cavity. The system achieves >23 dB of phase-noise improvement (at a 100 kHz offset) relative to a driving low-cost voltage-controlled oscillator (VCO) by exploiting BLs' optomechanical noise suppression and common-mode noise rejection via photomixing the two BLs. Uniform frequency stepping is ensured by locking each step to discrete, equally spaced cavity resonances, generating X-band SF waveforms spanning 1.31 GHz (8.16–9.47 GHz). The demonstrated system represents a path toward high-performance photonic radar with low phase noise and uniform frequency stepping—capabilities critical for resolving both weak and strong reflectors across a wide dynamic range for emerging applications, from healthcare monitoring to space exploration.

\begin{figure*}[ht!]
 \centering{
 \includegraphics{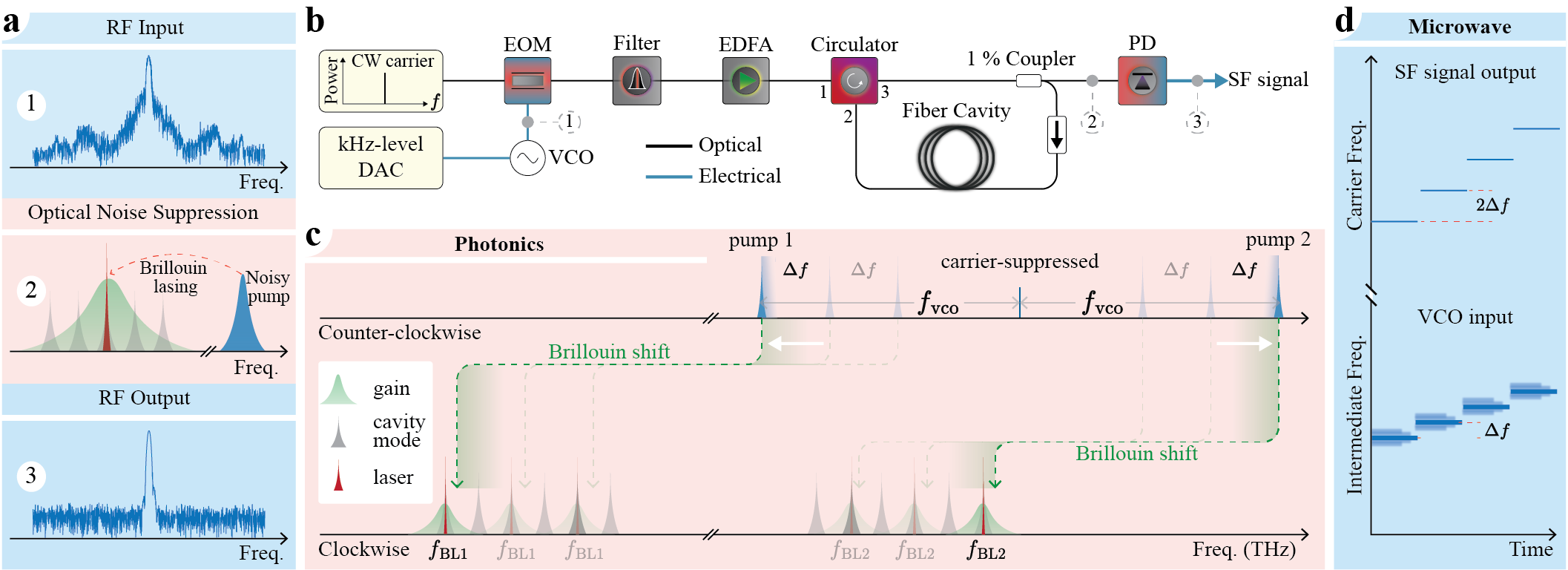}
            }
  \caption{
  \textbf{Principle of stepped-frequency (SF) radar signal generation enhanced by dual Brillouin lasers.} (\textbf{a}) Illustration of a microwave photonic link that uses Brillouin lasers to suppress the phase noise of an RF input signal. (\textbf{b}) System schematic. A voltage-controlled oscillator (VCO) drives an electro-optic modulator (EOM) to generate optical sidebands. Two sidebands are selected using an optical fiber. An erbium-doped fiber amplifier provides gain to the two sidebands, which are injected into a fiber cavity via an optical circulator. PD: photodetector; SF: stepped-frequency. (\textbf{c}) Optical frequency domain illustration of bi-chromatic pumping (pump 1 and pump 2) propagating in the counter-clockwise direction, and the corresponding dual Brillouin lasers propagating in the clockwise direction in a fiber ring cavity. The two pumps are generated from the first-order electro-optic modulation sidebands of a carrier, driven by an RF VCO at frequency $f_{\text{VCO}}$. Each pump induces a Stokes gain profile (green); when the gain exceeds the cavity loss, dual Brillouin lasers (red) are initiated. Optical spectra are shown at three successive time steps $t_{1}$, $t_{2}$, and $t_{3}$, corresponding to the first three VCO frequency increments of the SF waveform. (\textbf{d}) Time-frequency domain illustration of the input VCO waveform with high phase noise and the resulting output RF SF waveform with lower phase noise. 
  }
  \label{fig1}
\end{figure*}

\Fref{fig1} illustrates the operating principle of the demonstrated radar signal generator: an MWP link based on dual BLs in a fiber cavity. The system simultaneously suppresses the VCO's phase noise through Brillouin lasing (\fref{fig1}\textbf{a}), ensures uniform frequency stepping by pinning each lasing oscillation to a discrete cavity resonance equally spaced by the free spectral range (FSR), and increases the output bandwidth through photonics-based RF frequency multiplication. A VCO-generated driving signal—inherently with low bandwidth and high phase noise (\fref{fig1}\textbf{a})—is first transferred to the optical domain via electro-optic (EO) modulation (\fref{fig1}\textbf{c}). The EO modulator generates two first-order optical sidebands that carry the VCO's phase noise, serving as dual Brillouin pumps (pump 1 and pump 2) injected into a fiber ring cavity via an optical circulator (\fref{fig1}\textbf{b}) \cite{Secondini2023, lucas2023dynamic}. These pumps induce two Stokes signals in the counterpropagating direction (clockwise in \fref{fig1}\textbf{c}), each shifted downward in frequency by the Brillouin frequency shift. When the Stokes gain exceeds the round-trip cavity loss, the two Stokes modes start to oscillate as BLs. The phase noise of the resulting BLs is significantly reduced, owing to the combination of a short phonon lifetime and a long photon lifetime in the high-Q cavity, suppressing the transfer of pump phase noise to the laser emission \cite{gundavarapu2019sub}. Photomixing the two BLs then generates a low-phase-noise microwave signal. Additionally, because the two BLs share a common cavity, optical-path-length fluctuations are suppressed via common-mode noise rejection, further reducing phase noise at high offset frequencies \cite{geng2007dual, li2013microwave}.

The SF waveform is generated by incrementing the digital-to-analog converter (DAC) output voltage to step-tune the VCO frequency (\fref{fig1}\textbf{a}), which, in turn, step-tunes the optical sidebands acting as Brillouin pumps. Each frequency step successively excites dual Brillouin lasing oscillations at equally spaced cavity resonances (\fref{fig1}\textbf{c}), ensuring uniform frequency stepping across the waveform. Photonic RF frequency multiplication is achieved upon photomixing the two BLs; here, first-order sideband selection yields RF frequency doubling. As a result, a low-noise, bandwidth-broadened SF waveform with uniform frequency steps defined by the cavity resonance grid is generated (\fref{fig1}\textbf{d}).

\begin{figure*}[ht!]
 \centering{
 \includegraphics{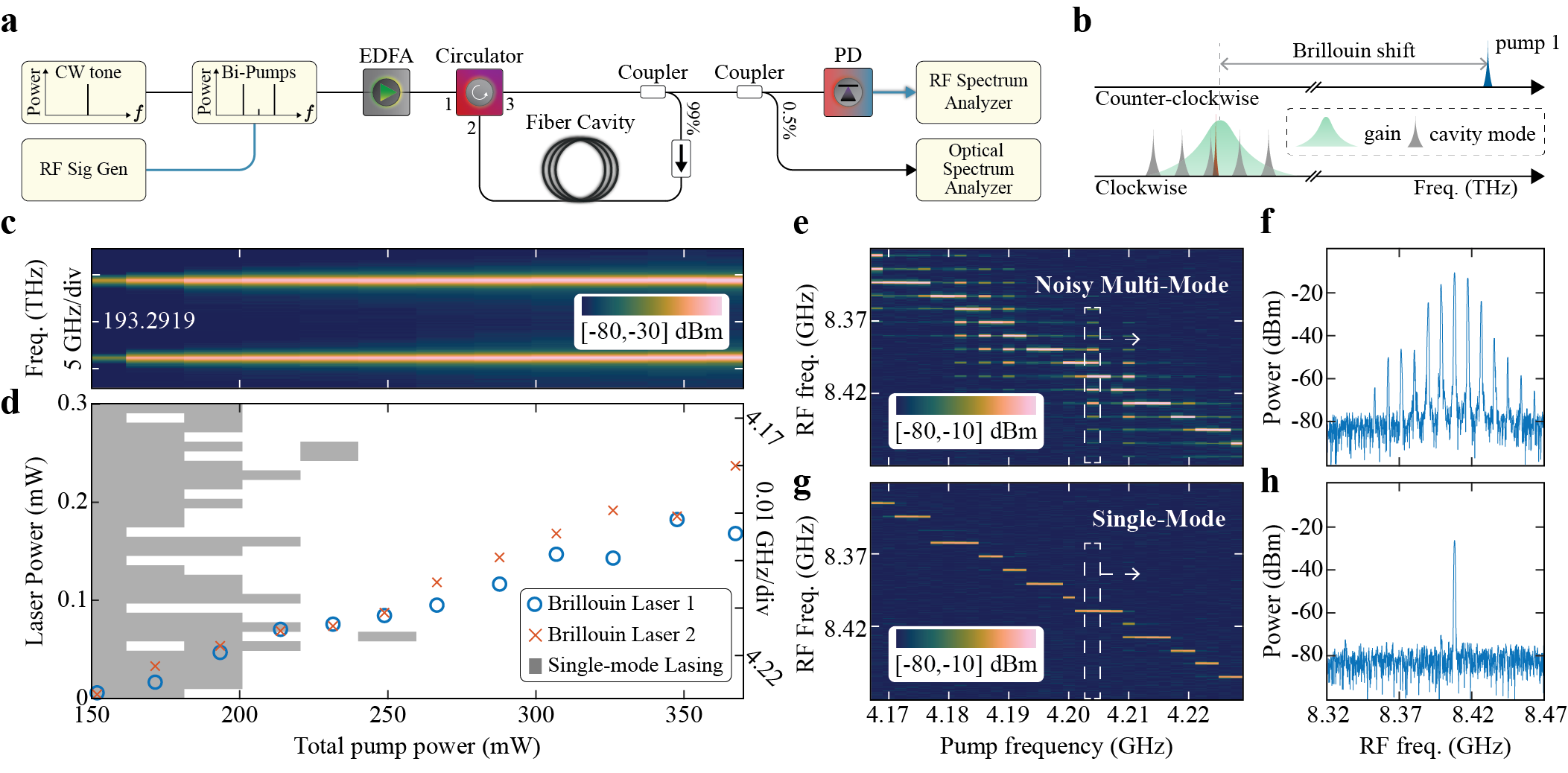}
            }
  \caption{\textbf{Brillouin laser emission regimes characterization.} 
(\textbf{a}) Schematic of the dual-pumped SF radar signal generator. The system contains a 22m-long nonreciprocal polarization-maintaining (PM) fiber cavity.
(\textbf{b}) Frequency-domain representation of a single pump (pump 1) relative to the cavity’s resonances, the Brillouin gain spectrum, and the resulting laser.
(\textbf{c}) Optical spectra of the two Brillouin lasers as a function of total pump power (sum of both pumps).
(\textbf{d}) Laser threshold measurement. The gray shows the system's single-mode lasing regime as a function of the RF signal generator’s frequency.
(\textbf{e}) RF spectrum of the beating signal observed in the multi-mode region at a high pump power of 370 mW.
(\textbf{f}) Example electronic spectrum of multi-mode lasers beating.
(\textbf{g}) RF spectrum of the beating signal observed in the single-mode region at a pump power of 152 mW.
(\textbf{h}) Example electronic spectrum of single-mode lasers beating.
  }
  \label{fig2}
\end{figure*}

The experimental setup is shown in \fref{fig2}\textbf{a}. Before characterizing radar performance, we map the Brillouin lasing regimes as a function of pump power and pump-cavity frequency detuning to identify the single-mode operating window and avoid the multi-mode regime, where phase noise degrades (see Supplementary). Thus, we sweep the pump laser frequency across several cavity free spectral ranges (FSRs) at different pump powers to map the lasing behavior.
% For this characterization, an RF signal generator drives a carrier-suppressed EO modulator to precisely control the frequency spacing between the two pumps, thereby avoiding the VCO's frequency drift due to environmental perturbations.
For this characterization, an RF signal generator drives a carrier-suppressed EO modulator to precisely control the frequency spacing between the two pumps. The first-order optical sidebands after the EO modulation are selected using an optical bandpass filter. The two pump lasers are amplified by an erbium-doped fiber amplifier (EDFA) and injected into a high-Q fiber cavity (see Table 1 in Method) to initiate Brillouin lasing. The pump power is controlled via the EDFA gain and monitored with an optical power meter, while the spectra of the two BLs are recorded simultaneously with an optical spectrum analyzer (OSA) and an electronic spectrum analyzer (ESA), in the optical and electrical domains, respectively. 

The two Brillouin laser spectra recorded across a range of total pump powers are shown in \fref{fig2}\textbf{c}. The extracted peak powers (\fref{fig2}\textbf{d}) show that neither laser is sustained when the total pump power decreases below 152 mW, indicating a per-pump lasing threshold of approximately 76 mW, assuming equal pump powers. At each pump power, we sweep the pump frequency across more than 6 cavity FSRs (by sweeping the signal generator from 4.168 GHz to 4.228 GHz) in 2 MHz steps to study the lasing stability. From the RF spectra, we map the single-mode lasing region as a function of pump offset frequency and pump power, shown in \fref{fig2}\textbf{d}. These results align with previous studies \cite{Lecoeuche1996} that the system tolerates larger pump-cavity detuning when operating near the lasing threshold. For instance, both lasers remain in the single-mode region across the full detuning range at a total pump power of 152 mW (\fref{fig2}\textbf{g} and \textbf{h}), whereas they operate exclusively in the multi-mode region at powers above 250 mW (\fref{fig2}\textbf{e} and \textbf{f}).  The maximum pump power is estimated to be 85.7 mW to avoid multi-mode operation, based on calculations of the gain of the adjacent modes (see Supplementary). However, the SF waveform (\fref{fig2}\textbf{e} and \textbf{g}) exhibits non-uniform frequency steps due to frequency drift between the carrier laser and the fiber cavity, which introduces mode-hopping.

\begin{figure*}[ht!]
 \centering{
 \includegraphics{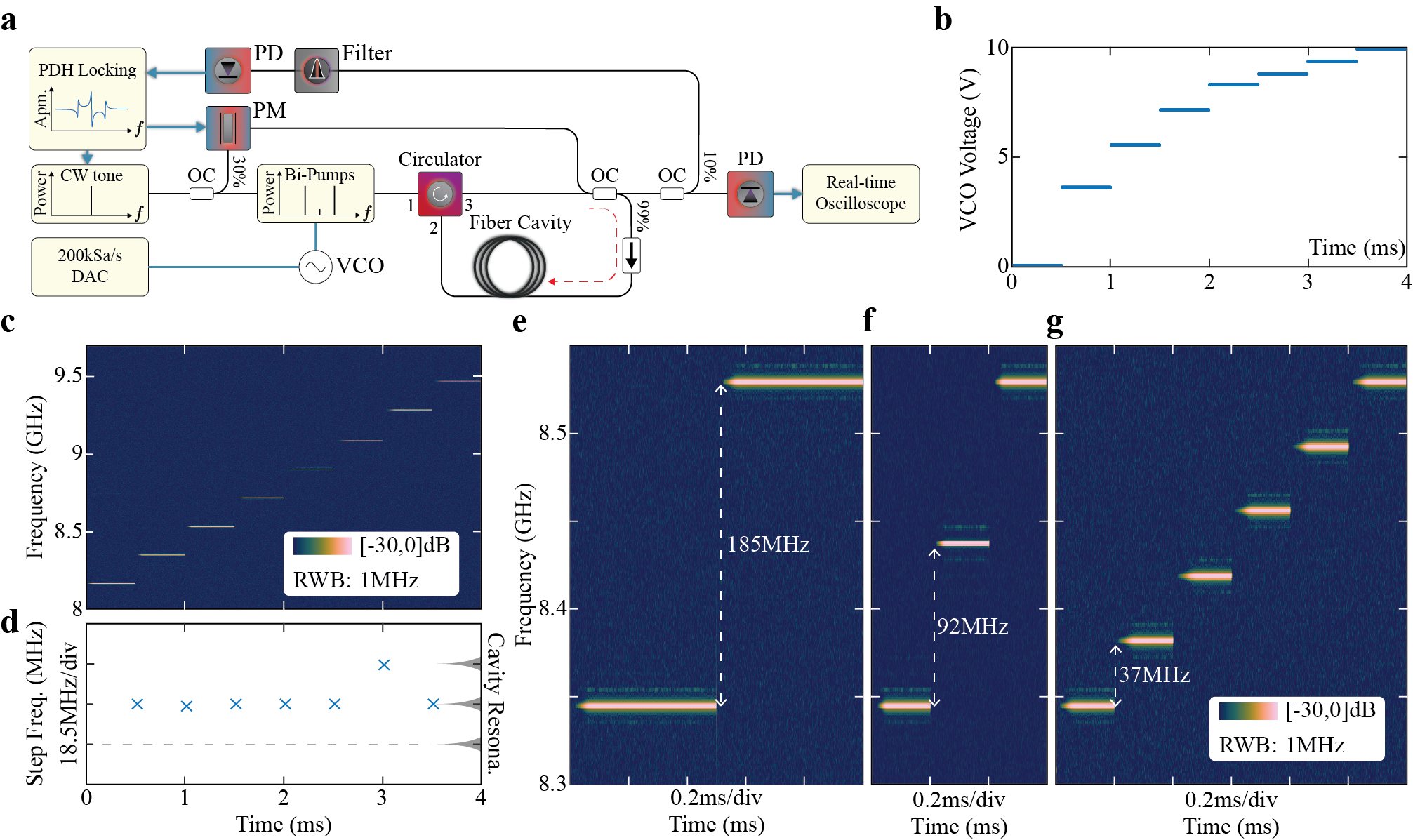}
            }
  \caption{\textbf{SF waveform generation using a VCO and a 200 kSa/s DAC.} (\textbf{a}) Schematic of the dual-pumped SF radar signal generator. Pound–Drever–Hall (PDH) locking is employed to stabilize cavity mode hopping.
  (\textbf{b}) Output voltage of the DAC, including a pre-distorted voltage profile used to linearise the time–frequency response of the VCO.
  (\textbf{c}) Time-frequency plot of the generated SF radar waveform with a frequency step of 185 MHz.
  (\textbf{d}) Measured step-frequency relative to the theoretical values of a perfectly linear SF waveform. The sixth step deviates by two FSRs due to the lack of temperature stabilization in the system.
  (\textbf{e-g}) Tunable SF waveforms with different frequency steps obtained by adjusting the DAC output voltage.
  }
  \label{fig3}
\end{figure*}

To overcome mode-hopping and generate an SF waveform with uniform frequency steps, we employ the Pound–Drever–Hall (PDH) locking technique \cite{black2001introduction} to lock the optical carrier to the fiber cavity (\fref{fig3}\textbf{a}). PDH locking eliminates the relative frequency drift between the laser and the fiber cavity, ensuring on-resonance pumping by aligning the peak Brillouin gain with the cavity resonance and suppressing uncontrolled mode-hopping that would otherwise produce non-uniform frequency steps. To achieve fast frequency stepping, the RF signal generator is then replaced with a low-cost VCO capable of rapid frequency hopping. The VCO frequency response is linearized using a pre-distorted voltage waveform (\fref{fig3}\textbf{b}) generated by a low-speed DAC operating at 200 kSa/s. By applying a short-time Fourier transform (STFT) to the system's output in the time domain (recorded by a real-time oscilloscope), we observe uniform frequency steps (\fref{fig3}\textbf{c}), enabled by PDH locking, which enforces lasing on a fixed, equally spaced grid of cavity modes. The SF waveform spans 8.16–9.47 GHz, with 7 frequency steps separated by 185 MHz and a 500 µs dwell time, yielding a total bandwidth of 1.31 GHz. The peak frequency extracted at each step confirms the uniform step spacing, as shown in \fref{fig3}\textbf{d}. A single outlier is observed at the sixth step, where the frequency hop is 18.5 MHz larger than the nominal spacing, corresponding to the offset of one cavity mode. This deviation is primarily caused by VCO temperature drift; thus, the VCO sweep nonlinearity has changed from the previous calibration and can be mitigated through temperature stabilization or feedback control of the VCO driving voltage. Smaller frequency steps can be easily achieved by adjusting the DAC output voltage to select cavity modes with smaller frequency spacing.  For example, step sizes of 92.5 MHz and 37 MHz are demonstrated in \fref{fig3}\textbf{f} and \fref{fig3}\textbf{g}, respectively. This tunability is essential for SF radar systems, as smaller frequency steps extend the unambiguous detection range (see Supplementary), a key performance metric in SF radar.

\begin{figure*}[ht!]
 \centering{
 \includegraphics{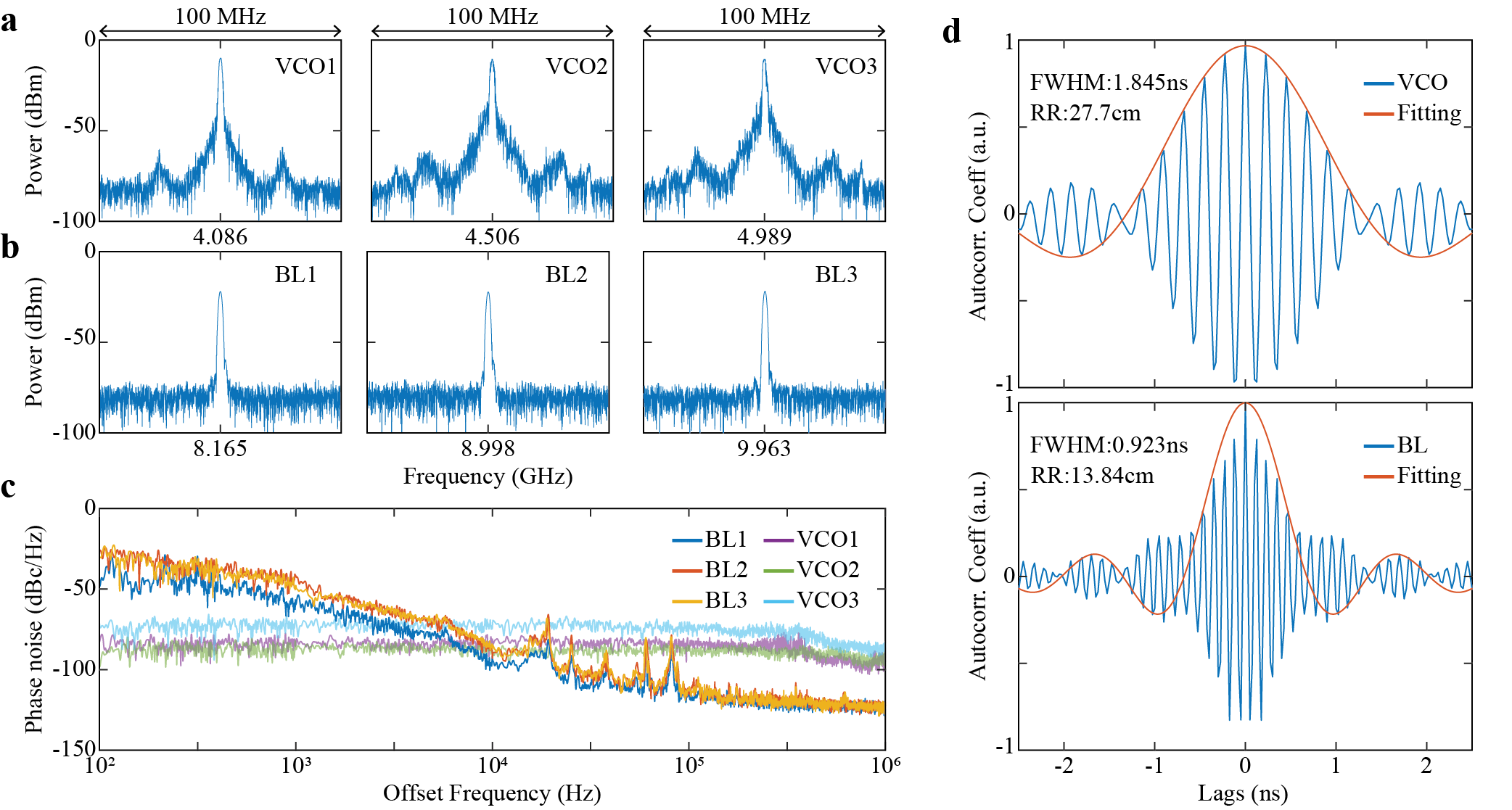}
            }
  \caption{\textbf{Performance characterization of the generated SF waveform.}
  (\textbf{a}) Measured RF spectrum of the free-running VCO, illustrating the high noise floor of the source.
  (\textbf{b}) RF spectrum of the system’s output, demonstrating low-noise signal generation with a suppressed noise pedestal.
  (\textbf{c}) Phase noise comparison between the VCO input signal and the low-noise system output, measured using an electronic spectrum analyzer. Measurements are taken at three VCO frequencies, 4.1, 4.5, and 5.0 GHz (labeled VCO1, VCO2, and VCO3, respectively), along with their corresponding system output frequencies of 8.2, 9.0, and 10.0 GHz (labeled BL1, BL2, and BL3, respectively).
  (\textbf{d}) Autocorrelation of the time-domain waveforms for both the VCO and the Brillouin system. The narrowed autocorrelation peak of the system's output confirms the improved range resolution, from 27.7cm to 13.84 cm, achieved through frequency-doubling, i.e., the beating of the first-order sidebands.
  }
  \label{fig4}
\end{figure*}

We next characterize the system's phase noise suppression capability using the ESA, measuring the RF spectrum and phase noise at three representative frequencies (approximately 8.2, 9.0, and 10 GHz) via the ESA's built-in phase noise analyzer.  As shown in \fref{fig4}\textbf{a} and \textbf{b}, the VCO noise pedestals are strongly suppressed, thanks to the denoising process of Brillouin lasing and common-mode noise cancellation during photodetection. The phase noise measurements reveal a consistent improvement of more than 23 dB at a 100 kHz offset across all three frequencies (\fref{fig4}\textbf{c}). It is worth noting that the output RF waveform is dominated by $f^{-3}$ noise component (see Supplementary), primarily attributed to the pump's relative intensity noise (RIN) and environmental perturbations modeled as thermorefractive and photothermal phase noise \cite{dallyn2022thermal}. Both noise contributions can be mitigated through active intensity stabilization of the pump laser \cite{li2014low} and thermal stabilization of the fiber cavity, the latter of which has been shown to significantly reduce photothermal noise. Finally, we demonstrate the SF radar's range resolution by autocorrelating the time-domain signal recorded by a high-speed oscilloscope (see Supplementary). As shown in \fref{fig4}\textbf{d}, the autocorrelation peak narrows from 27.7 cm (VCO alone) to 13.8 cm (system output)—a factor of two reduction consistent with the twofold bandwidth increase introduced by the photonic frequency-doubling process. These results (\fref{fig4}) confirm that the system successfully generates a low-noise X-band SF radar waveform with doubled RF bandwidth, validating its potential for high-performance radar sensing.

\noindent\textbf{Discussion} High spectral purity and low phase noise are critical for maximizing the dynamic range of radar systems, particularly in airborne and satellite-based sensing \cite{cooper2020compact}. A reduced noise floor enables the detection of weak or slow-moving targets that would otherwise be obscured by the noise pedestals of strong nearby reflectors \cite{ghelfi2014fully, kittlaus2021low}. In satellite-based meteorology, for instance, high spectral purity enables the resolution of weak reflections from low-altitude clouds despite ground reflections that are orders of magnitude stronger. By suppressing the phase-noise pedestal that typically masks these subtle atmospheric signals, the low-noise SF radar architecture demonstrated here provides a high-performance, cost-effective pathway for next-generation remote sensing technologies.

The demonstrated system is also promising for operation at higher RF carrier frequencies, extending into the millimeter-wave (mmWave) and terahertz (THz) regimes, enabling ultra-wideband radar signal generation and superior range resolution. By selecting higher-order harmonics to increase the frequency multiplication factor, the system can achieve an $N$-fold increase in the input bandwidth, yielding an $N$-fold finer range resolution. For instance, replacing the intensity modulator with a dual-parallel Mach-Zehnder modulator (DPMZM) enables direct selection of second-order harmonics as Brillouin pumps via simple bias control \cite{zhang2017photonics}. This frequency quadrupling approach would quadruple the input RF bandwidth, potentially achieving a range resolution of 7 cm based on the current experimental setup. Furthermore, W-band (75-110 GHz) waveforms can be generated using a narrowband VCO at higher carrier frequencies (e.g., the HMC8364), enabling output sweeps from 96 GHz to 106 GHz and achieving millimeter-level range resolution.

The system also holds significant potential for full integration onto heterogeneous photonic platforms. The bulk modulator can be replaced with a high-performance lithium niobate (LN) \cite{feng2024integrated, zhu2025integrated, hu2025integrated} or lithium tantalate-based platform \cite{zhu20258, niels2026high, su2026lithium, zhang2025ultrabroadband}, while optical gain can be achieved with semiconductor optical amplifiers (SOAs) or erbium-doped silicon nitride (SiN) waveguides \cite{liu2022photonic, liu2024fully, xinru2026nc}. Additionally, meter-long ultra-high-$Q$ integrated ring resonators are now available on platforms such as SiN \cite{kai2025, 2025heim} and germanium-doped silica (Ge-silica) \cite{chen2026towards}. These platforms can achieve fiber-like, ultra-high $Q$-factors in the hundreds of millions with MHz-level FSR, simultaneously ensuring high cavity enhancement for superior noise suppression and a reduced frequency step to extend the radar’s unambiguous range \cite{zhang2024photonic}. These platforms can also incorporate Bragg grating structures \cite{russell2025brillouin} to inhibit higher-order Brillouin cascading, thereby mitigating laser linewidth broadening \cite{behunin2018fundamental} and ensuring low noise.

% However, it is worth noting that moving to integrated ring resonators with reciprocal cavities introduces a specific noise challenge: Brillouin laser cascading enhances laser noise \cite{behunin2018fundamental}, broadening the linewidth and potentially worsening the radar’s performance. This issue can be mitigated, for instance, by incorporating Bragg grating structures within the waveguide to create a photonic bandgap \cite{russell2025brillouin}. Such structures can selectively inhibit higher-order Brillouin cascading, thus ensuring the low-noise feature of the SF radar signal in a fully integrated architecture. 

In conclusion, we have demonstrated a photonic SF radar architecture based on dual BLs in a fiber cavity, simultaneously achieving low phase noise, wide bandwidth, and high time-frequency linearity—key performance metrics that existing photonic approaches struggle to achieve. The system requires only a low-cost VCO and a low-speed DAC for the RF input, providing a simple, cost-effective architecture for simultaneous phase noise suppression, bandwidth broadening, and uniform frequency stepping. The demonstrated system generates X-band SF waveforms with >23 dB of phase noise reduction at a 100 kHz offset and the potential for millimeter-level range resolution by leveraging a higher frequency multiplication factor. Compatible with heterogeneous photonic integration, the architecture offers a promising pathway toward next-generation remote sensing platforms, particularly for SWaP-constrained airborne and satellite-based applications.

\section*{Methods}

\noindent The system employs a continuous-wave (CW) laser (Teraxion, NLL1550). Owing to the limited tunability of this laser module, direct Pound–Drever–Hall (PDH) locking is not feasible in the demonstrated experiment. Instead, a tunable CW optical tone is synthesized using a voltage-controlled oscillator (VCO), an EO modulator, and an optical bandpass filter to select the first-order modulation sideband at the higher frequency. This sideband serves as the CW tone, as shown in \fref{fig2}\textbf{a} and \fref{fig3}\textbf{a}. 

A typical PDH locking system was implemented by using a servo controller (Sacher Lasertechnik, LB2001 and LB2005). The feedback voltage is sent to the VCO (Mini-Circuits, ZX95-5540C-S+) to lock its frequency to the fiber cavity resonance. The dual Brillouin pumps are generated using an intensity modulator (Exail, MXER-LN-20) driven by a second VCO (Mini-Circuits, ZX95-5400-S+). The non-reciprocal Brillouin fiber cavity consists of a polarization-maintaining (PM) circulator, a PM isolator, and a 20 m-long PM fiber (Coherent, PM1550B-XP), yielding a total cavity length of about 22.1 m. 

\input{table1}
Optical spectra are measured using an optical spectrum analyser (Finisar, WaveAnalyzer 1500S). RF time-domain signals are recorded with a real-time oscilloscope (Agilent, DSOX96204Q) with a sampling speed of 80 GSa/s, and frequency-domain measurements are performed using an electronic spectrum analyzer (Agilent, E4448A). The real-time oscilloscope recorded time-domain waveforms, which were later used to generate the short-time Fourier transform plots shown in \fref{fig3}. The STFT was calculated using a 1 $\mu$s sliding window with 50\% overlap and a 1 MHz FFT resolution.

\noindent \textbf{Acknowledgments} This work was supported in part by the Australian Research Council Centre of Excellence Project under Grant CE230100006 and in part by the University of Sydney DVCR Strategic Research Impact Fund under Grant DVCR POC 30-2024. We acknowledge the discussion with Eric Magi, Moritz Merklein, and Wenle Weng.

\noindent \textbf{Data Availability} The data that support the plots within this paper will be available upon publication.

\noindent \textbf{Author Contributions} Z.Z. and B.J.E. conceived the project; Z.Z. designed the system structure; Z.Z. conducted the experiments; Z.Z., R.L.R, and C.K.L. performed the data analysis; Z.Z. wrote the manuscript with contributions from R.L.R, B.J.E., and C.K.L

\noindent \textbf{Competing Interests Statement} The authors declare no competing interests.

% \bibliography{library}
\bibliography{references}

\end{document}

\title{Brillouin-Enhanced Photonic Stepped-Frequency Radar: Supplemental Document}
%--------------------------1---------------------------
\author{Ziqian Zhang}
\email{ziqian.zhang@sydney.edu.au}
\affiliation{Institute of Photonics and Optical Science (IPOS), School of Physics, The University of Sydney, NSW 2006, Australia}
\affiliation{The University of Sydney Nano Institute (Sydney Nano), The University of Sydney, NSW 2006, Australia}
\affiliation{ARC Centre of Excellence in Optical Microcombs for Breakthrough Science (COMBS), NSW 2006, Australia}
%--------------------------2---------------------------
\author{Ryan L. Russell}
\affiliation{Institute of Photonics and Optical Science (IPOS), School of Physics, The University of Sydney, NSW 2006, Australia}
\affiliation{The University of Sydney Nano Institute (Sydney Nano), The University of Sydney, NSW 2006, Australia}
\affiliation{ARC Centre of Excellence in Optical Microcombs for Breakthrough Science (COMBS), NSW 2006, Australia}
%--------------------------3---------------------------
\author{Choon Kong Lai}
\affiliation{Institute of Photonics and Optical Science (IPOS), School of Physics, The University of Sydney, NSW 2006, Australia}
\affiliation{The University of Sydney Nano Institute (Sydney Nano), The University of Sydney, NSW 2006, Australia}
\affiliation{ARC Centre of Excellence in Optical Microcombs for Breakthrough Science (COMBS), NSW 2006, Australia}
%--------------------------5---------------------------
\author{Benjamin J. Eggleton}
\email{benjamin.eggleton@sydney.edu.au}
\affiliation{Institute of Photonics and Optical Science (IPOS), School of Physics, The University of Sydney, NSW 2006, Australia}
\affiliation{The University of Sydney Nano Institute (Sydney Nano), The University of Sydney, NSW 2006, Australia}
\affiliation{ARC Centre of Excellence in Optical Microcombs for Breakthrough Science (COMBS), NSW 2006, Australia}
%------------------------------------------------------
% \date{\today}
\maketitle

\begin{adjustwidth}{1cm}{2cm}
    \tableofcontents
\end{adjustwidth}
\newpage
\section{SF Radar Ranging Principle}

To explain the ranging principle, we consider an SF radar illuminating a stationary plane reflector, which is treated as a point-like target for simplicity. In this configuration, the SF radar transmits a signal consisting of $N$ discrete frequency steps, \( n = 1, ..., N \), each separated by a frequency increment of  \( \Delta f \) and lasting for a duration of $T_{\text{step}} $.

The demodulation process examines the phase difference between the transmitted and received signals at each frequency step: 
\begin{equation}
\omega(n)= 2\pi\Delta f \tau \cdot n,
\end{equation} where $\tau$ is the roundtrip delay. For the following discussions and this equation, we consider only a single sample per step, indexed by $n$. $\omega(n)$ shows linear phase progression at a rate of $2\pi\Delta f \tau $ for a stationary target. Because the range information ($\tau$) is encoded in the signal phase, which is periodic with $2\pi$, a coherent receiver that obtains in-phase (I) and quadrature (Q) components is required to retrieve the target range.

Then, an inverse Fourier transform (IFT) is applied to the complex signal, formed by I, $\cos(\omega)$, and Q, $\sin(\omega)$, components, to extract the target’s range information.
\begin{equation}
\text{IFT} \left\{ \cos{\omega(n)} + i \sin{\omega(n)} \right\}
\end{equation} The IFT transforms the frequency-domain data into the time domain, revealing the target's range information. This contrasts with the Fourier transform, which performs the reverse operation by converting time-domain signals into their frequency-domain representations. 

\section{SF Radar Ambiguity Range}

Ambiguity range is a key metric in radar systems, which measures the range over which the true range of targets can be determined without aliasing. Ambiguity arises from the repetition rate of a pulsed radar system; it occurs when the radar echo arrives at the system beyond its original period. In a non-pulsed, CW radar system, for instance, the SF radar system, the ambiguity range is no longer related to the pulse repetition rate, since the waveform is transmitted continuously. SF radar ambiguity arises from another periodicity in the waveform, i.e., the equally spaced frequency steps. A coherent receiver is commonly used in SF radar systems, where the transmitted and received SF signals are mixed, and the mixing product at a lower frequency near DC is sampled to extract phase information, which can be expressed as:
\begin{equation}
    \phi_{n} = 2\pi \cdot n \Delta f \cdot \frac{2d}{c}
\end{equation}where $n = 1,2,...N$, $N \Delta f$ is the total bandwidth, $\Delta f$ is the step frequency introduced by the mode hopping of the Brillouin laser, $d$ is the distance of the target, and $c$ is the propagation speed of the RF signal. Consider a relative distance change to the original location $d$ as $\Delta d$, then the corresponding phase changes can be written as
\begin{equation}
    \Delta \phi_{n} = 2\pi \cdot n \Delta f \cdot \frac{2\Delta d}{c},
    \label{eq2}
\end{equation} which has a periodical phase change of $2\pi$. If the relative phase changes are more than the period ($2\pi$), ambiguities are introduced ($\Delta \phi > 2\pi$). Thus, by substituting the maximum phase changes, $2\pi$ into \Eref{eq2} and moving $\Delta d$ to the left side, the unambiguous range can be expressed as
\begin{equation}
    \Delta d_{\text{amb}} = \frac{c}{2 \Delta f}.
    \label{eq3} 
\end{equation}
In the paper, we presented three frequency steps: 185 MHz, 92 MHz, and 37 MHz, corresponding to ambiguity ranges of 0.81, 1.63, and 4.05 meters, respectively. The longest ambiguity range is set by twice the cavity FSR due to the symmetry of the EO sidebands. Two approaches can be used here to extend the system ambiguity range: 1) use a longer cavity to reduce FSR, and 2) configure the VCO to hop pseudo-randomly in frequency \cite{axelsson2007analysis}.  

\section{Radar Range Resolution}

Range resolution in radar refers to the system's capability to distinguish targets located closely in range. For wideband radar systems utilizing signals, such as LFM and SF, the resolution is proportional to the inverse of the signal's bandwidth, defined as follows \cite{Ozdemir2021}:
\begin{equation}
    r=  \frac{c}{2\text{B}}
\end{equation}
where $c$ is the propagation speed of the signal, and B is the total synthesized bandwidth. Therefore, increasing the signal’s bandwidth can improve the radar's range resolution.

\section{Phase Noise and Measurements}

Phase noise refers to random fluctuations in the phase of a waveform. In the time domain, random phase fluctuations cause deviations from a perfect sine wave, known as jitter. Three typical units of measure are used to quantify phase noise: phase instability, the noise energy, and the short-term frequency instability. We presented the results based on the ESA-measured noise energy, defined as $\mathcal{L}(f_m)$, which is the ratio of the power in the one-sided noise to the total signal power at an offset $f_m$. The ESA (Agilent, E4448A) has an amplitude accuracy of $\pm 0.29$ dB.

\begin{figure*}[ht!]
 \centering{
 \includegraphics{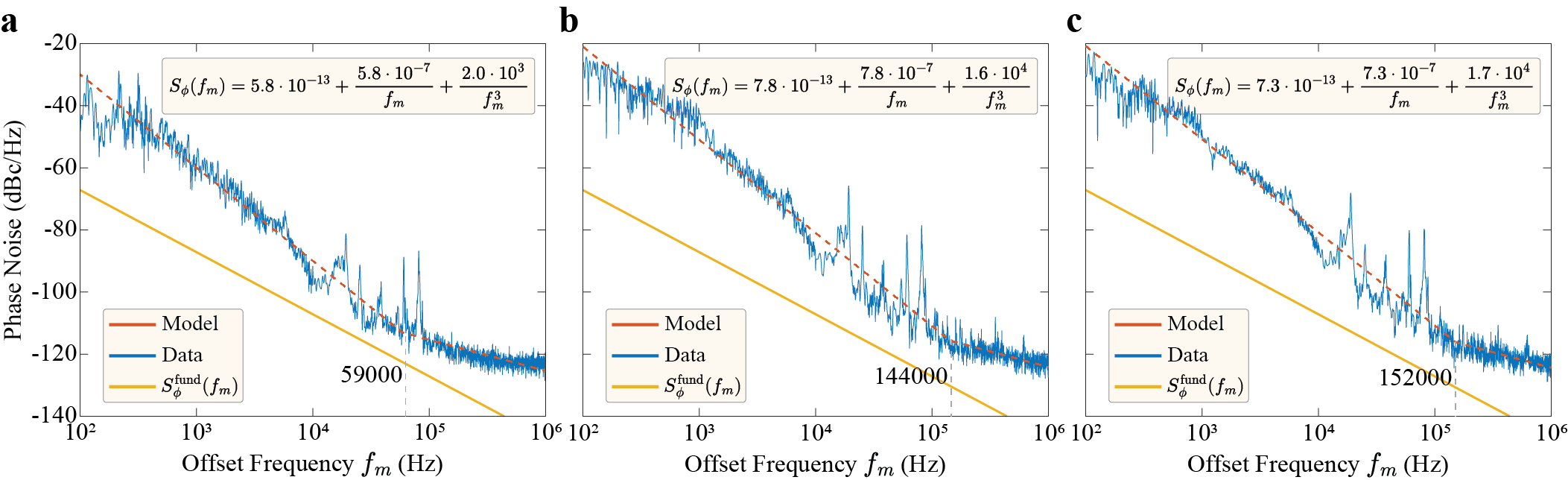}
            }
  \caption{\textbf{RF Phase Noise Fitting.} (\textbf{a}), (\textbf{b}) and (\textbf{c}) correspond to the fitting of the data of BL1, BL2, and BL3, respectively, shown in Fig. 4\textbf{c}.
  }
  \label{figS2}
\end{figure*}

We now fit the experimental data to the power-law (algebraic phase noise) model to give insights into the noise processes. The model is given in the phase instability:
\begin{equation}
   S_{\phi}(f_m) = k_{0} + \frac{k_{1}}{f_m} + \frac{k_{2}}{f_m^{2}} + \frac{k_{3}}{f_m^{3}} + \frac{k_{4}}{f_m^{4}}
\end{equation} where $k_{0}$ represents the white phase noise floor, $k_{1}$ represents flicker noise of phase, and $k_{2}$, $k_{3}$, and $k_{4}$ correspond to the contribution of white frequency noise, flicker frequency noise, and random walk frequency noise, respectively. For the following fittings, we use the $ 2 \mathcal{L}(f_m) = S_{\phi}(f_m) $ relationship under the small angle ($\Delta \phi$) approximation \cite{packard1985phase}. The fitting results are shown in \fref{figS2}, revealing that the phase noise behavior of this RF source is dominated by the $k_{3}$ (flicker) noise component.

There are four dominant sources of phase noise in Brillouin lasers: 1) fundamental noise intrinsic to the physics of SBS lasing, $S_{\phi}^{\mathrm{fund}}$, 2) transferred frequency noise from the pump laser, $S_{\phi}^{\text{trans}}$, 3) photothermal noise, $S_{\phi}^{\mathrm{PT}}$, and 4) thermorefractive noise, $S_{\phi}^{\mathrm{TR}}$. The model is given as \cite{dallyn2022thermal}:
\begin{equation}
S_{\phi}(f_m)=S_{\phi}^{\mathrm{fund}}(f_m)
+S_{\phi}^{\text{trans}}(f_m)
+S_{\phi}^{\mathrm{PT}}(f_m)
+S_{\phi}^{\mathrm{TR}}(f_m).
\end{equation} Here, we adopted the well-established model (modified Schawlow-Townes) to describe the fundamental noise \cite{suh2017phonon},
\begin{equation}
    S_{\phi}^{\text{fund}}(f_m) =\frac{\hbar f_{\text{BL}}^3n_T}{2P Q_T Q_E}\cdot \frac{1}{f_m^{2}}.
        \label{sfund}
\end{equation} where $\hbar$ is the Planck constant, $f_{\text{BL}}$ is the Brillouin laser's frequency, $n_T$ is the number of thermal quanta in the mechanical field at the Brillouin shift frequency ($\Omega$), $P$ is the laser output power, $Q_T$ is the loaded Q-factor, and $Q_E$ represnts the external Q-factor. The thermal phonon number is computed using the Bose-Einstein model at a room temperature of 298 K. Based on Table S1, we calculate the fundamental noise $S_{\phi}^{\text{fund}}$, which is shown in \fref{figS2}. The fundamental noise floor ($f_{m}^{-2}$) is obscured by the dominant $f_{m}^{-3}$ noise component. In the demonstrated system, this dominant noise is primarily attributed to two sources: 1) the pump's relative intensity noise (RIN), and 2) environmental perturbations, modeled as thermorefractive and photothermal phase noise \cite{dallyn2022thermal}. RIN can be mitigated by actively stabilizing the pump laser's intensity \cite{li2014low}. Separately, thermal stabilization of the fiber cavity has been shown to significantly reduce the impact of photothermal noise \cite{loh2019ultra}.

\input{table2}

The transferred frequency noise is given by \cite{debut2000linewidth},
\begin{equation}
    S_{\phi}^{\text{trans}}(f_m) =\left(\frac{\gamma}{\gamma+\Gamma_{\text{B}}}\right)^2 S_{\phi}^{\text{pump}}(f_m),
\end{equation} where $\gamma = 2\pi f_{\text{BL}}/Q_T$, and $S_{\phi}^{\text{pump}}(f_m)$ represents the frequency noise of the pump. Based on Table S1, the calculated optical phase noise suppression applied to the pump is approximately -37.5 dB.

\section{Laser Stability chart}

Here, we describe the simulation of the laser's stability chart. As in coherent Brillouin-Kerr comb generation, mode hopping and multi-mode lasing are the two main limiting factors, including the detuning range required to access different comb states. Both mode hopping and multi-mode lasing are related to two key parameters: the pump power level and the ratio of the Brillouin gain bandwidth to the cavity FSR. We calculate the stability chart of a Brillouin laser based on gain and the adjacent modes in the cavity, neglecting other factors, such as optical mode competition over a common acoustic field. The model tends to underestimate the power threshold, $P_{\text{max}}$, that separates single-mode and multi-mode lasing due to mode competition and hysteresis, which tend to extend the single-mode operation of an initially lasing mode \cite{lucas2023dynamic}. 
\begin{equation}
\frac{P_{\max }}{P_{t h}^j} \approx 1+\left[2 \frac{D_1}{\Gamma_B}\right]^2,
\end{equation} where
\begin{equation}
P_{t h}^j=\frac{A_{\mathrm{eff}} \kappa}{g_{B} 2 c} \approx \frac{A_{\mathrm{eff}}}{g_{B}} \frac{D_{\mathfrak{1}}[1-\eta(1-\theta)]}{2 \pi c},
\end{equation} where $P_{t h}^j$ is the lasing threshold, $D_1$ is the cavity FSR, $\Gamma_B$ is the Brillouin gain bandwidth, $\eta$ is the discrte cavity loss, and $\theta$ is the out-coupling ratio. It is worth mentioning that the lasing threshold ($P_{t h}^j$) assumes the gain is on resonance. 

\begin{figure*}[ht!]
 \centering{
 \includegraphics{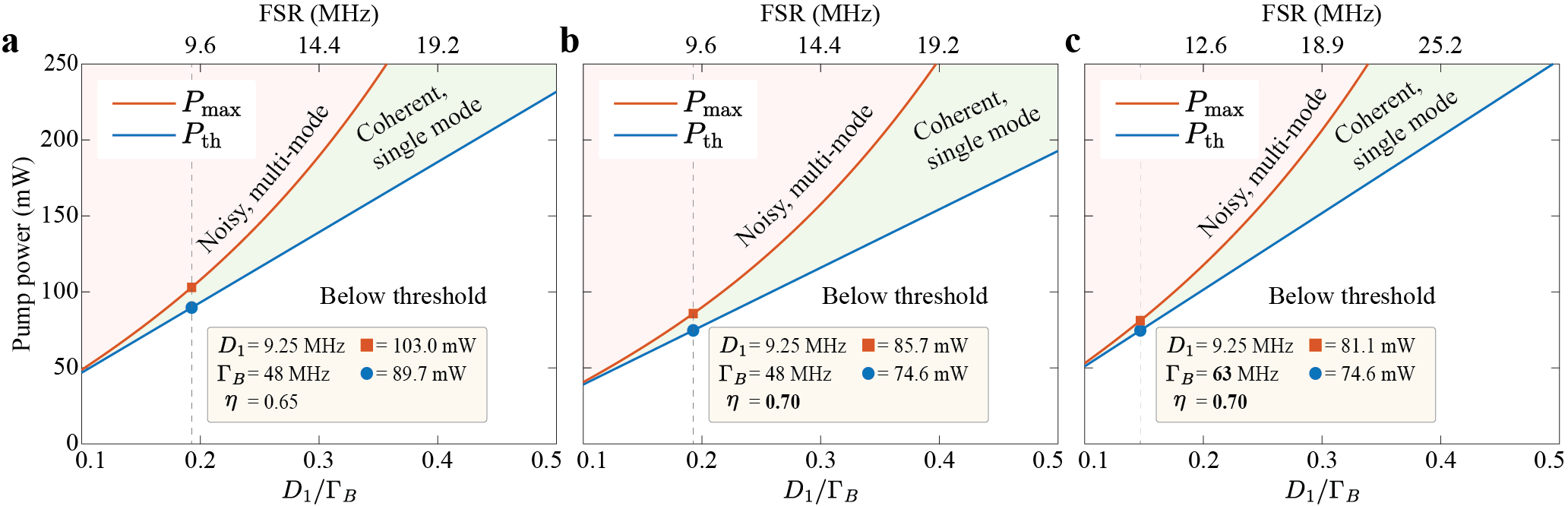}
            }
  \caption{\textbf{Simplified stability chart} A function of cavity FSR, $D_{1}$, relative to the Brillouin gain bandwidth, $\Gamma_{B}$.
  }
  \label{figS3}
\end{figure*}

In the above calculation, we use the experimentally measured value of $\eta = 0.65$ (-1.87 dB loss) and obtain a Brillouin-laser threshold of 90 mW (\fref{figS3}\textbf{a}), which is slightly higher than the experimental value of approximately 75 mW. One possible reason for this discrepancy is that the intra-cavity losses are lower than the individual component measurements indicate. The calculated threshold is 75 mW (\fref{figS3}\textbf{b}) when we set $\eta = 0.7$ (-1.55 dB loss), which aligns with the experiment. A narrower Brillouin gain bandwidth ($\Gamma_{\text{B}}$) is preferred for this architecture, as it reduces the number of cavity modes that fall within the gain profile (\fref{figS3}\textbf{b} and \textbf{c}). This effectively minimizes mode competition, extends the pump's stable dynamic range, and avoids the noisy, multi-mode lasing.

\section{Mode Pulling}

Brillouin laser mode (frequency) pulling is well-studied. The Brillouin gain introduces a dispersive phase shift inside the cavity. The dispersion modifies the round-trip phase of the intra-cavity field and "pulls"  the laser oscillation towards the line center of the Brillouin gain \cite{Lecoeuche1996, Li2012}. The following formula is based on the assumption of a "cold cavity", and the mode-pulling slope is given by 

\begin{equation}
\frac{\partial \omega_{\text {Stokes }}}{\partial \delta \Omega}=\frac{1}{1+\Gamma_B / \kappa} = 1.36 \times 10 ^{-2}, 
\end{equation} where $\Gamma_B = 48$ MHz and $\kappa = 660$ kHz are the gain bandwidth and linewidth, respectively. For a narrow-linewidth cavity, the frequency shift caused by mode pulling is at the kHz level; for instance, $\partial \omega_{\text{Stokes}} = 1.36$ kHz when $\partial \delta \Omega = 1$ MHz. This means that a smaller frequency instability from the VCO (e.g., due to environmental factors) will cause a negligible change in the linearity of the frequency step. 

\section{Comparative time-domain analysis}

\begin{figure*}[ht!]
 \centering{
 \includegraphics{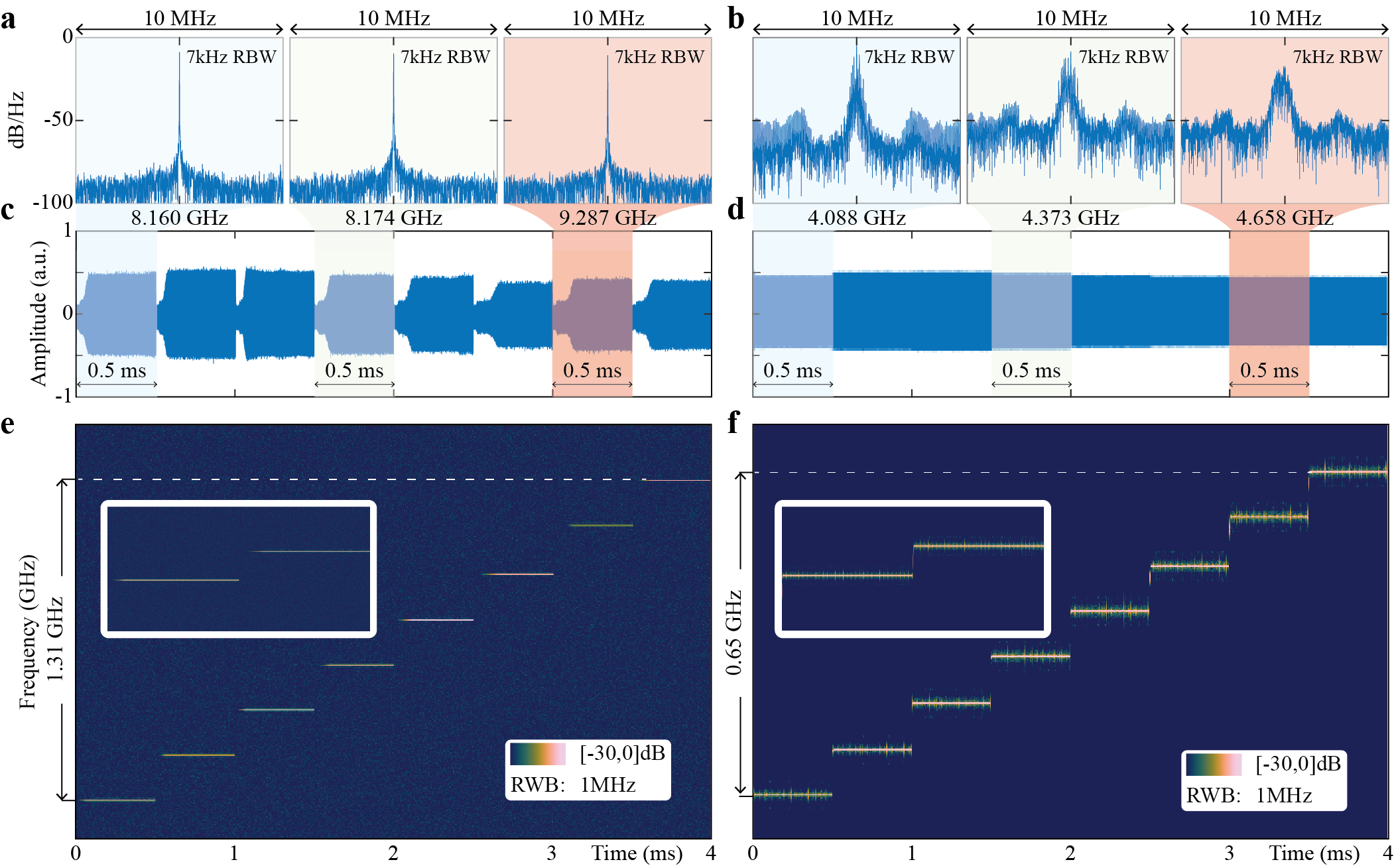}
            }
  \caption{\textbf{Comparative time-domain and time-frequency analysis.} (\textbf{a}, \textbf{b}) Calculated power spectral densities (PSDs) for the system output and the raw VCO input, respectively. The spectra are calculated from time-domain data at three distinct frequency steps. (\textbf{c}, \textbf{d}) Time-domain waveforms captured via a high-speed oscilloscope for the noise-suppressed output (\textbf{c}) and the raw VCO input (\textbf{d}). The transient build-up behavior during frequency hopping is visible in (\textbf{c}). (\textbf{e}, \textbf{f}) Short-time Fourier transform (STFT) of the corresponding time-domain data.
  }
  \label{figS1}
\end{figure*}

Here, we provide a comparative time-domain analysis of the VCO input and the system's output. \fref{figS1}\textbf{c} and \fref{figS1}\textbf{d} record the raw time-domain waveforms for the system and the VCO, respectively. The corresponding power spectral densities (PSDs), calculated via the periodogram method, are presented in \fref{figS1}\textbf{a} and \fref{figS1}\textbf{b}. These were derived from a 0.5 ms time-domain segment comprising 20 million data points, using a Kaiser window ($\beta = 38$) to achieve a 7.014 kHz resolution bandwidth. A comparison of these spectra reveals that the raw VCO (\fref{figS1}\textbf{b}) exhibits significantly higher noise levels within the dwell time of each step compared to the noise-suppressed system output (\ref{figS1}\textbf{a}).

To investigate the temporal dynamics of these signals beyond the 0.5-ms average, we performed a Short-Time Fourier Transform (STFT), as shown in \fref{figS1}\text{e} and \fref{figS1}\text{f}. The STFT was calculated using a 1 $\mu$s sliding window with 50\% overlap and a 1 MHz FFT resolution. The magnified views in \fref{figS1}\text{e} and \fref{figS1}\text{f} highlight the inherent frequency and noise instabilities of the free-running VCO, while simultaneously confirming the high spectral purity and temporal stability maintained by the Brillouin-based generation system.

It should be noted that the analyses in this section differ from the results in Fig. 4, as the latter were obtained in the frequency domain using an electronic spectrum analyzer without active step-tuning of the VCO. Furthermore, the transient build-up time observed for each frequency step in \fref{figS1}\textbf{c} is attributed to frequency instabilities and imperfect pre-distortion in the VCO drive signal. Specifically, the preset tuning voltages failed to instantaneously align the pump frequency with the cavity resonance, thereby prolonging the time required for the Brillouin laser to reach a steady state. This effect could be mitigated in future implementations by thermally stabilizing the VCO and applying adaptive learning algorithms to optimize the frequency-tuning pre-distortion \cite{zhang2019}.

\section{Frequency Multipilier}
Phase noise refers to the frequency instability of a signal and is distinct from the noise figure (NF), which measures the added broadband thermal noise. Phase noise is quantified as $\mathcal{L}(f_m)$. When frequency multiplication occurs, it increases phase noise by a factor of \( 20 \log_{10}(n) \), where \(n\) is the multiplication factor \cite{Secondini2023}. Therefore, frequency doubling ($n=2$) and quadrupling ($n=4$) incur a noise penalty of 6.02 dB and 12.04 dB, respectively. It is worth noting that the demonstrated system not only avoided the 6.02 dB noise penalty but also reduced it by more than 23 dB. 

\bibliography{references.bib}

%% file: table1.tex
\begin{table}[ht!]
\centering
\label{tab1}
\caption{Cavity parameters}
\begin{tblr}{
  hline{1,9} = {-}{0.08em},
  hline{2} = {-}{},
}
Cavity Parameters &  & Unit\\
Gain bandwidth ($\Gamma_{\text{B}}$) & 48 & MHz\\
Total cavity length ($L$) & 22.1 & m\\
Roundtrip loss ($\eta$) & -1.87& dB\\
Coupling ratio ($\theta$) & 0.9  & \% \\
Loaded linewidth ($\kappa$) & 660 & kHz\\
Free spectral range & 9.25 & MHz\\
Loaded Q & 297 & M

\end{tblr}
\end{table}

%% file: table2.tex
\begin{table}[ht!]
\centering
\label{tab2}
\caption{Cavity parameters}
\begin{tblr}{
  hline{1,10} = {-}{0.08em},
  hline{2} = {-}{},
}
Cavity Parameters &  & Unit\\
Gain bandwidth ($\Gamma_{\text{B}}/2\pi$) & 48 & MHz\\
Loaded linewidth ($\kappa$) & 660 & kHz\\
Loaded Q ($Q_{T}$) & 297 & M\\
External Q ($Q_{E}$) &14522 & M\\
Brillouin laser frequency ($f_{\text{BL}}$) & 193.278 & THz\\
Laser output power ($P$) & 0.013 & mW \\ 
Brillouin shift ($\Omega$) & 10.86 GHz & GHz \\
Room Temperature (T) & 298 & Kelvin
\end{tblr}
\end{table}